\documentclass[twocolumn,prd,showpacs,superscriptaddress,preprintnumbers]{revtex4}

\usepackage{graphicx,epsfig}
\usepackage{bm}
\usepackage{latexsym,amssymb,amsmath,float}
\usepackage{subfig,caption}
\usepackage[countmax]{subfloat}

\newcommand{\beq}[1]{\begin{equation}\label{#1}}
\newcommand{\eeq}{\end{equation}}
\newcommand{\bea}[1]{\begin{eqnarray} \label{#1}}
\newcommand{\eea}{\end{eqnarray}}
\newcommand{\ba}{\begin{array}}
\newcommand{\ea}{\end{array}}

\newcommand{\rf}[1]{(\ref{#1})}
\def\be{\begin{equation}}
\def\ee{\end{equation}}
\def\gs{\mathrel{
   \rlap{\raise 0.511ex \hbox{$>$}}{\lower 0.511ex \hbox{$\sim$}}}}
\def\ls{\mathrel{
   \rlap{\raise 0.511ex \hbox{$<$}}{\lower 0.511ex \hbox{$\sim$}}}}

\newcommand{\rarr}{\rightarrow}

\newcommand{\bad}{\begin{array}{ccc}}
\newcommand{\bav}{\begin{array}{cccc}}
\newcommand{\baf}{\begin{array}{ccccc}}

\def\U2{\underline{U\hspace{-.9mm}}\hspace{.9mm}}
\def\nubar{\bar\nu}
\def\nue{{\nu_{e}}}
\def\numu{{\nu_{\mu}}}
\def\nutau{{\nu_{\tau}}}

\newcommand{\nua}{\nu_\alpha}

\newcommand{\nub}{\nu_\beta}

\newcommand{\enot}{{\not e}}

\begin{document}

\title{Cosmic Neutrino Flavor Ratios with Broken $\nu_\mu$--$\nu_\tau$ Symmetry}

\author{Lingjun Fu}
\email[Electronic mail: ]{lingjun.fu@vanderbilt.edu}
\affiliation{Department of Physics and Astronomy,
Vanderbilt University, Nashville, TN 37235, USA}

\author{Chiu Man Ho}
\email[Electronic mail: ]{chiuman.ho@vanderbilt.edu}
\affiliation{Department of Physics and Astronomy,
Vanderbilt University, Nashville, TN 37235, USA}

\author{Thomas J. Weiler}
\email[Electronic mail: ]{tom.weiler@vanderbilt.edu}
\affiliation{Department of Physics and Astronomy,
Vanderbilt University, Nashville, TN 37235, USA}

\date{\today}

\begin{abstract}
Reactor neutrino experiments have now observed a nonzero value for $\theta_{13}$ at $\,5\sigma$,
and global fits to data imply a nonzero value above $\,10\sigma$.
Nonzero values for $\theta_{13}$ and/or $\theta_{32}$-$\frac{\pi}{4}$ break a $\nu_\mu$-$\nu_\tau$ symmetry,
which has qualitative as well as quantitative implications for the time-evolution of neutrino flavors.
In particular, the large-distance flavor evolution matrix,
non-invertible with $\nu_\mu$-$\nu_\tau$ symmetry, is now invertible.
This means that measurements of neutrino flavor ratios at Earth can now be
inverted to directly reveal the flavor ratios injected at cosmically distant sources.
With the updated values of the three neutrino mixing angles, we obtain the inverted large-distance evolution matrix and
use it to derive several phenomenological relations between the injection flavor ratios and the observable ratios at Earth.
Taking the three popular injection models as examples, we also exhibit the shift of
Earthly observed flavor ratios from the corresponding values in the case with $\nu_\mu$-$\nu_\tau$ symmetry.
\end{abstract}


\maketitle

\section{Introduction}
\label{sec:intro}
It is well-known that a statistical average over a neutrino ensemble from cosmic distances eliminates the
quantum-mechanical phase $\phi_{jk}\equiv L(m_j^2-m_k^2)/2E$ between states,
leaving a relatively simple result for neutrino-flavor evolution.
The evolution $\nua\rarr\nub$, with $\alpha$ and $\beta$ any elements of the three-flavor set
$\{e,\ \mu,\ \tau\}$, is
described by the propagation matrix $P$ whose positive definite elements are
\beq{Pab}
P_{\alpha\beta}= \sum_j  | U_{\alpha j} |^2\, | U_{\beta j} |^2\,.
\eeq
The physics behind this formula is that the correct basis for particle propagation is the mass
basis, because the particle propagator in field theory is an analytic function
with poles at mass values;
we sum over the unobserved mass states labeled by $j$, and we weight each such mass state by its
classical probability $|U_{\alpha j}|^2$ to overlap with the flavor $\alpha$ produced at the source,
times its classical probability $|U_{\beta j}|^2$ to overlap with the flavor $\beta$ detected at Earth.
The sum on $j=1,2,3$ is over the three active neutrino states.
(Astrophysical distances are so much larger than relevant oscillation lengths that
subtleties in the definition of cosmic distance~\cite{WagWei} are not important here.)

Phase-averaging restores $CP$-invariance, and so the matrix $P$
describes both neutrino and anti-neutrino propagation.
Furthermore, $CP$-invariance, according to the $CPT$-theorem, implies also $T$-invariance,
and so the matrix $P$ is symmetric, $P_{\alpha\beta} = P_{\beta\alpha}$.
Explicitly, one has
\beq{}
P =
 \left(
\ba{ccc}
\sum_j |U_{e j}|^2\,|U_{e j}|^2 & \sum_j |U_{e j}|^2\,|U_{\mu j}|^2 & \sum_j |U_{e j}|^2\,|U_{\tau j} |^2 \\
 & & \\
 \cdots                        & \sum_j |U_{\mu j}|^2\,|U_{\mu j}|^2 & \sum_j |U_{\mu j}|^2\,|U_{\tau j} |^2 \\
 & & \\
 \cdots                        & \cdots                             & \sum_j |U_{\tau j}|^2\,|U_{\tau j} |^2
\ea
\right)
\eeq
Thus, the flavor ratio unit-vector injected at the source,
${\vec W} \equiv (W_e, W_\mu, W_\tau)$
is measured at Earth to be ${\vec w}\equiv (w_e, w_\mu, w_\tau)$, where
\beq{FlavorProp1}
{\vec w} = P\,{\vec W}\,.
\eeq

If $P$ is an invertible matrix (i.e., has a nonvanishing determinant),
then the inverse equation
\beq{FlavorProp2}
{\vec W} = P^{-1}\,{\vec w}
\eeq
allows one to input the neutrino flavor ratios observed at Earth
to obtain the flavor ratios dynamically injected at the cosmic source.
At present, the only observed sources of extra-terrestrial neutrinos are from
the Sun and SN1987A. The hope is that neutrino telescopes, recently
deployed~\cite{IceCube,Antares},
or soon to be deployed~\cite{KM3net},
will begin to observe neutrinos from more distant sources.
When an ensemble of neutrino events are collected, track-topologies will allow
one to glean the ratios of flavors arriving at Earth~\cite{MeasuringFlavor}.
Very recently, the IceCube experiment has announced what is likely the first observation
of high-energy extra-galactic neutrinos,
two showering events characteristic of $\nue$'s
(or $\bar\nu_e$'s, since the experiments cannot distinguish $\nu$ from $\nubar$),
with energies $\sim$PeV.
It appears that the era of neutrino astrophysics is suddenly upon us.

On the terrestrial neutrino experimental front,
neutrino mixing data were consistent with $\numu$-$\nutau$ symmetry,
defined operationally as $|U_{\mu j}| = |U_{\tau j}|$, until very recently.
The popular example of such a $\numu$-$\nutau$ symmetric model is
the TriBimaximal (TBM) model~\cite{TBM},
which has the following classical probability and flavor propagation matrices:
\beq{UsquaredTBM}
|U_{\alpha j}|^2 = \frac{1}{6}
\left(
\ba{ccc}
 4 & 2 & 0 \\
 1 & 2 & 3 \\
 1 & 2 & 3
\ea
\right)
=
\left(
\ba{ccc}
 0.667 & 0.333 & 0     \\
 0.167 & 0.333 & 0.500 \\
  0.167 & 0.333 & 0.500
\ea
\right)\,,
\eeq
and
\beq{P-TBM}
P_{\rm TBM}=\frac{1}{18}
\left(
\ba{ccc}
10 & 4 & 4  \\
 4  & 7 & 7 \\
 4  & 7 & 7 \\
\ea
\right)
=
\left(
\ba{ccc}
0.55 & 0.22 & 0.22 \\
0.22 & 0.39 & 0.39 \\
0.22 & 0.39 & 0.39 \\
\ea
\right)\,.
\eeq

With exact $\numu$-$\nutau$ symmetry, the second and third rows of the $|U|^2$ and $P$ matrices are identical,
by definition.  Accordingly, the determinant of matrix $P$ vanishes, and the matrix is not invertible.
Thus, prior theoretical work, which attempted to relate flavor ratios at Earth to the ratios injected at the cosmic sources,
guessed at the source ratios and
and evolved the guesses forward with Eq.~\rf{FlavorProp1} to obtain the observable ratios at Earthly detectors.
Popular guesses are the pion decay-chain flavor-ratios $\frac{1}{3}$(1:2:0), the $\beta$-beam ratios (1:0:0)~\cite{beta-beam},
and the incomplete pion decay-chain or quenched $\mu$-decay ratios (0:1:0)~\cite{MeszarosWaxman}.
We will examine flavor evolution in these three injection models in some detail below.
A thoughtful overview of neutrino injection models is given in~\cite{Hummer:2010ai}.

More satisfying would be to approach the study with observed flavor ratios,
and evolve them backwards via Eq.~\rf{FlavorProp2} to obtain as directly as possible
the astrophysical quantities of interest, namely the flavor ratios injected at the sources.
Recent neutrino data~\cite{theta13} from nuclear reactors reveals that the $\numu$-$\nutau$ symmetry is broken.
Hence, the determinant of $P$ is no longer vanishing, and the inverse flavor propagation matrix $P^{-1}$ is calculable.

The first purpose of this paper is to provide this inverse propagation matrix.
As the second purpose of this investigation,
we will draw various phenomenological inferences for three-flavor neutrino astrophysics.
For example, we will plot the movement of Earthly flavor ratios away from their
TriBiMaximal values, for the three most popular cosmic-source flavor models;
present a relation among flavor ratios at Earth that determines whether tau neutrino's are injected at the source;
derive a general formula for the injection flavor ratio at the source in terms of the
observable ratio of readily measurable track and shower events at Earth;
and derive bounds on the possible flavor ratios to be observed on Earth,
as implied by the measured mixing angles.
If observations at Earth were to violate these latter bounds,
then some physics other than flavor mixing via phase-averaged vacuum oscillations would be at play.
Examples of new physics could be neutrino decay~\cite{nuDK},
or oscillations into new states such as sterile neutrinos or pseudo-Dirac states~\cite{pDirac}.

\section{Two-Component Flavor -- The TBM Example}
\label{sec:TBM}
It is instructive to see how flavor bounds may be derived in the simple case of TBM mixing.
The $\numu$-$\nutau$ symmetry tells us that $\numu$ and $\nutau$ will arrive in equal numbers,
regardless of the flavor distribution at the source.
Thus, there are but two relevant flavor ratios at Earth, $w_e$ and $w_{\enot}$, with
$w_\enot$ equally split between $\numu$ and $\nutau$, and $w_e + w_\enot = 1$.
We may obtain the propagation matrix in this $(e,\enot)$ basis by adding the identical $\numu$ and $\nutau$
rows in Eq.~\rf{P-TBM}, and omitting the now redundant third column.  One gets
\beq{TBMeff}
P_{\rm TBM}^{\rm eff} = \frac{1}{9}
\left(
\ba{cc}
 5 & 2 \\
 4 & 7
\ea
\right)\,.
\eeq
The propagation equation
\beq{twobytwo}
\left(
\ba{c}
w_e \\
w_\enot
\ea
\right)
=
P_{\rm TBM}^{\rm eff}
\left(
\ba{c}
W_e \\
W_\enot
\ea
\right)
\eeq
is linear, so the flavor extremes at Earth are found by inputting into
Eq.~\rf{twobytwo} the pure flavor vectors ${\vec W}=(1,0)$ and ${\vec W}=(0,1)$ at the source.
The results are trivially $2/9 \le w_e \le 5/9$, correlated with
$7/9 \ge w_\enot \ge 4/9$.
Such is the allowed region of flavor space at Earth (${\vec w}$) in the TBM model.
%

Proceeding further, the determinant of $P_{\rm TBM}^{\rm eff}$ is nonzero,
and so this matrix is invertible.
The inverse matrix is
\beq{inverseTBMeff}
(P_{\rm TBM}^{\rm eff})^{-1} =  \frac{1}{3}
\left(
\ba{rr}
  7 & -2 \\
  -4 &  5
\ea
\right),
%
%
\left(
\ba{c}
W_e \\
W_\enot
\ea
\right)
=
(P_{\rm TBM}^{\rm eff})^{-1}
\left(
\ba{c}
w_e \\
w_\enot
\ea
\right)\,.
\eeq
%
From Eq.~\rf{inverseTBMeff} we derive an interesting expression relating flavor ratios
at the source to the same ratio observed at Earth:
\beq{R(r)}
\frac{W_e}{W_{\not e}} = \frac{7\left(\frac{w_e}{w_{\not e}}\right)-2}{5-4\left(\frac{w_e}{w_{\not e}}\right)}\,.
\eeq
Neutrino telescopes are particularly adept at distinguishing muon tracks due to $\numu$ interactions,
from showering events due to $\nue$ and $\nutau$ interactions.
We may bin together the latter events as $w_{sh}\equiv w_e +w_\tau$.
Then, inputting the TBM relations $w_e=w_{sh}-w_\mu$ and $e_{\not e}=2w_\mu$ into Eq.~\rf{R(r)},
we get an alternative expression of the same relation,
\beq{R(r)2}
\frac{W_e}{W_{\not e}} = \frac{7-11\left(\frac{w_\mu}{w_{sh}}\right)}{14\left(\frac{w_\mu}{w_{sh}}\right)-4}\,.
\eeq
These TBM relations, Eqs.\rf{R(r)} and \rf{R(r)2}, hold for any injection model.
If either LHS were output, its value would discriminate among injection models.

Theory strongly suggests that $\nutau$ production at the source is very small~\cite{Lipari,PRWei}
(i.e., $W_\tau$ is zero or nearly so, or equivalently, that $W_\enot\approx W_\mu$).
In hand with this assumption,
the inverse matrix $(P^{\rm eff}_{\rm TBM})^{-1}$ of Eq.~\rf{inverseTBMeff}
provides a complete reconstruction of the flavor ratios at the source
in terms of flavor ratios observed at Earth, for the TBM-mixing model.

\section{Broken \boldmath{$\numu$-$\nutau$} Symmetry and Three-Component Flavor}
\label{threenus}
With the observation of $\theta_{13}$ far from zero, we learn that $\numu$-$\nutau$ symmetry is likely broken,
and an analysis of the full three-component vectors $\vec w$ and $\vec W$
of Eqs.~\rf{FlavorProp1} and \rf{FlavorProp2} is warranted.
Let us begin with a short review of the evidence for broken  $\numu$-$\nutau$ symmetry.

Following \cite{PDG}, the three-neutrino mixing matrix is conventionally parameterized by three planar rotations,
analogous to the three Euler angles of classical mechanics,
and a purely quantum-mechanical Dirac phase~$\delta$~\footnote{
The simplest relations between the three angles and three mixing moduli are
$|U_{e3}|^2 = s^2_{13}$, $|U_{e2}|^2 = c^2_{13} s^2_{12}$, and $|U_{\mu 3}|^2 = c^2_{13} s^2_{32}$.
Also simple are $|U_{e1}|^2 = c^2_{13} c^2_{12}$ and $|U_{\tau 3}|^2 = c^2_{13} c^2_{32}$,
but these two are not independent of the other three relations.
There is a fourth independent $|U_{\alpha j}|^2$, which must be chosen from the set
$\{|U_{\mu 1}|^2,|U_{\mu 2}|^2,|U_{\tau 1}|^2,|U_{\tau 2}|^2\}$.
The independence of four $|U_{\alpha j}|^2$'s corresponds to the independent three mixing angles
plus one Dirac phase $\delta$.}:
\beq{UPDG}
U_{PMNS}\equiv R_{32}(\theta_{32})\,R_{13}(\theta_{13},\delta)\,R_{12}(\theta_{12})\,.
\eeq
The double argument of $R_{13}$ is a reminder that by convention, the quantum mechanical Dirac phase
appears in the off-diagonal elements of the rotation in the 13-plane:
\beq{13rotation}
R_{13} =
\left(
\ba{ll}
\ \ \ \cos\theta_{13} & \ \sin\theta_{13}e^{-i\delta}\\
-\sin\theta_{13}e^{+i\delta} & \ \cos\theta_{13}
\ea
\right)\,.
\eeq
The range for the angles is $[0,\pi/2 ]$, while the range for the phase is $[-\pi,+\pi]$,
or equivalently, $\cos\delta \in [-1,+1]$.

In terms of the three angles and the single phase,
one finds that the conditions $|U_{\mu j}| = |U_{\tau j}|$ for $\numu$-$\nutau$ symmetry require that
(i) $\theta_{32}=\pi/4$ and (ii) $\sin(2\theta_{12})\sin\theta_{13}\cos\delta = 0$.

Choices of angles/phase which satisfy (ii) above, necessary to uphold the $\numu$-$\nutau$ symmetry,
are~\cite{MarfWhis}:\,\,
\begin{itemize}
  \item Case (a):\,\, $\theta_{13}=0$
(TBM mixing is a special subcase of $\theta_{13}=0$, wherein $\sin\theta_{12}$ is set to $1/\sqrt{3}$);
  \item Case (b):\,\, $\theta_{12}=0$ or $\pi/2$;
  \item Case (c):\,\, $\cos\delta=0$, i.e.\ $\delta=\pm\pi/2$.
\end{itemize}
The recent spectacular evidence that $\theta_{13}\sim 9^\circ$ is not only nonzero,
but many~$\sigma$ from zero \cite{theta13},
rules out case (a).
The value of $\theta_{12}$ is inferred from experiment to be far from either zero or $\pi/2$,
which rules out case~(b).
(In addition, the matter effect responsible for suppression of $\nue$ from the sun requires
$|U_{e1} |> |U_{e2}| > 0$, which also rules out case~(b).)

Finally, we are left to discuss case~(c).
There is little experimental constraint on $\delta$, for it occurs (\,Eq.~\rf{13rotation}\,)
with $\sin\theta_{13}$ as a prefactor, and until this year
$\sin\theta_{13}$ was consistent with zero.
Recent experimental evidence~\cite{theta32} suggests that $\theta_{32}$ is not equal to
the maximal-mixing value $\pi/4$.
Thus, case (c) is mildly disfavored by data.
So it appears that $\numu$-$\nutau$ symmetry is likely broken.
If so, then $P$ is an invertible matrix.
And even if $\theta_{32}$ were exactly equal $\pi/4$,
it still remains a possibilty that $\numu$-$\nutau$ symmetry is broken
by a value $\delta \ne \pm\pi/2$.
In the rest of this section we proceed to analyze 3-neutrino flavor propagation
in the now favored case of a (slightly) broken $\numu$-$\nutau$ symmetry.

\subsection{Three-Flavor Analysis}
\label{subsec:threenu}
The constraint $W_e+W_\mu+W_\tau=1$ reduces the three-dimensional $W_e,W_\mu,W_\tau$-space
to the physical triangle with corners at $(1,0,0)$, $(0,1,0)$, and $(0,0,1)$.
Because the relation in Eq.~\rf{FlavorProp1} is linear in ${\vec W}$, the extremes of ${\vec w}$
are obtained from the values of ${\vec W}$ at these corners.
The result is
\beq{w-maxmin}
\max/\min\{w_\alpha\}=\max/\min\{P_{\alpha e},P_{\alpha\mu},P_{\alpha\tau}\}\,.
\eeq
(The two-component analog of this result was presented below Eq.~\rf{twobytwo}.)

For the theoretical expectation that $\nutau$'s are not produced by the cosmic mechanisms,
$W_\tau = 0$ and the extremes of ${\vec w}$ are even simpler:
\beq{w-maxmin-notau}
\max/\min\{w_\alpha\}=\max/\min\{P_{\alpha e},P_{\alpha\mu}\}\,,\
({\rm no\ source\ }\nutau {\rm 's}) 
\eeq
These simply-stated results have profound meaning.
For example, a measurement of any $w_\alpha$ satisfying Eq.~\rf{w-maxmin} but not satisfying~\rf{w-maxmin-notau}
would establish that in fact $\nutau$'s are produced at cosmically-distant sources.
(Note that there is no two-component analog of the $W_\tau=0$ result,
because with the parameter $W_{\not e}$ does not distinguish among $W_\tau$ and $W_\mu$.)

Inversion of the symmetric $3\times 3$ matrix $P$ yields
the symmetric inverse matrix
\begin{widetext}
\beq{Pinverse1}
P^{-1} = \frac{1}{{\rm Det}(P)}
\left(
\ba{ccc}
(P_{\mu\mu} P_{\tau\tau} - P^2_{\mu\tau} ) &
(P_{e\tau} P_{\mu\tau} - P_{e\mu} P_{\tau\tau} ) &
(P_{e\mu} P_{\mu\tau} - P_{e\tau} P_{\mu\mu} ) \\
 & & \\
\cdots & 
(P_{ee} P_{\tau\tau} - P^2_{e\tau} ) &
(P_{e\tau} P_{e\mu} - P_{ee} P_{\mu\tau} )  \\
 & & \\
\cdots & 
\cdots & 
(P_{ee} P_{\mu\mu} - P^2_{e\mu} )
\ea
\right)\,.
\eeq
\end{widetext}
With this $P^{-1}$~matrix in hand, we may directly calculate the injection ratios
${\vec W} = P^{-1}{\vec w}$ once the Earthly ratios ${\vec w}$ are measured.
This matrix, along with the numerical values given below for its matrix elements,
are among our main results.

As mentioned earlier,
the inverse of $P$, namely
$P^{-1}_{\alpha\beta} = {\rm Det}^{-1}(P)\;(-1)^{\alpha+\beta}\times {\rm cofactor\ of\ element\ } P_{\alpha\beta}$,
does not exist for the TBM model, because ${\rm Det}(P_{\rm TBM})$ vanishes.
However, the sign-weighted cofactor matrix, $(-1)^{\alpha+\beta}\times {\rm cofactor\ of\ } (P_{\rm TBM})_{\alpha\beta}$,
does exist, and its form sheds light on what one can expect in the realistic three-neutrino case
for the relative values of matrix elements in  $P^{-1}$.
From the values of the TBM matrix in Eq.~\rf{P-TBM},
one can easily calculate the sign-weighted cofactor matrix, ``${\rm Det}(P)\,P^{-1}$'',
even for matrices with a vanishing determinant.
(The quotation marks here are merely a reminder that ``${\rm Det}(P)\,P^{-1}$''
is just a convenient label for the sign-weighted cofactor matrix.)
We get
\beq{TBMinverse}
``[ {\rm Det}(P_{\rm TBM})\,(P_{\rm TBM})^{-1} ]" = \frac{1}{6}
\left(
\ba{crr}
  0 & 0 & 0 \\
  0 & 1 & -1 \\
  0 & -1 & 1 \\
\ea
\right)\,,
\eeq
and we repeat for the reader that the right-hand side of this relation is calculated independently of ${\rm Det}(P)$.
Because the updated values of the mixing angles break the $\numu$-$\nutau$ symmetry implicit
in the TBM model by only a small amount,
we may expect that with updated values of mixing angles,
matrix element values similar to those of Eq.~\rf{TBMinverse} will result for ``${\rm Det}(P)\,P^{-1}$'',
and a texture similar to that in Eq.~\rf{TBMinverse} will result for the inverse matrix $P^{-1}$.

Of course, with small  $\numu$-$\nutau$ symmetry breaking,
${\rm Det}(P)$ will also be small, but as long as it is nonzero, the matrix $P$ is invertible.

It is instructive to see how the shift in ${\rm Det}(P)$ will happen.
${\rm Det}(P)$ is unchanged when we subtract the second row of $P$ from the bottom row,
and then subtract the second column of $P$ from the third column.
After a bit of algebra, the result is
\beq{DetP}
{\rm Det}(P) = \left|
\ba{ccc}
P_{ee} & P_{e\mu} & \sum_j |U_{ej}|^2 \Delta_j \\
 & & \\
P_{e\mu} & P_{\mu\mu} & \sum_j |U_{\mu j}|^2 \Delta_j \\
 & & \\
\sum_j |U_{ej}|^2 \Delta_j\ \ & \sum_j |U_{\mu j}|^2 \Delta_j & \sum_j \Delta_j \Delta_j \\
\ea
\right|
\eeq
where $\Delta_j\equiv |U_{\tau j}|^2-|U_{\mu j}|^2$ are the three parameters characterizing
the breaking of $\numu$-$\nutau$ symmetry~\cite{ZZXing}.
Only two of the three $\Delta_j$ are independent, since
$\sum_j \Delta_j = \sum_j |U_{\mu j}|^2 - \sum_j |U_{\tau j}|^2 = 1 - 1 = 0$.
The symmetry of $P$ allows the row subtraction and column subtraction in
${\rm Det}(P)$ to each introduce a factor of $\Delta_j$, leading to a determinant
that is of order~$(\Delta_j \Delta_k)$, very small.
In the form of Eq.~\rf{DetP}, the determinant is easily evaluated.
The result is
\begin{widetext}
\beq{DetP2}
{\rm Det}(P) =
\left( \sum_j \Delta_j\,\Delta_j \right)\,(P_{ee}\,P_{\mu\mu} - P^2_{e\mu}) +
  \sum_{j,k} \Delta_j\,\Delta_k\,
     \left( \,2\,P_{e\mu}\,|U_{ej}|^2\,|U_{\mu k}|^2 -P_{\mu\mu}\,|U_{ej}|^2 |U_{ek}|^2
      -P_{ee}\,|U_{\mu j}|^2\,|U_{\mu k}|^2\, \right)\,.
\eeq
\end{widetext}
%
In terms of the mixing angles and Dirac phase, we note that the $\numu$-$\nutau$
symmetry-breaking parameters have expressions
\bea{symbreaking}
\Delta_1 &=& c(2\theta_{32})\,(c^2_{12}\,s^2_{13}-s^2_{12})-s(2\theta_{12})\,s(2\theta_{32})\,s_{13}\cos\delta\,,\nonumber\\
\Delta_2 &=& c(2\theta_{32})\,(s^2_{12}\,s^2_{13}-c^2_{12})+s(2\theta_{12})\,s(2\theta_{32})\,s_{13}\cos\delta\,,\nonumber\\
\Delta_3 &=& c(2\theta_{32})\,c^2_{13}\,.
\eea

\subsection{Three-Flavor Numerics}
\label{subsec:Numerics}
Perturbative expansions about TBM values for neutrino flavor ratios and for the inference of the Dirac phase $\delta$
have been considered previously, both before~\cite{TriMin} the measurement of $\theta_{13}$, and after~\cite{Meloni}.
None of the prior work considers the inversion of  the propagation matrix.
In what follows, we will not appeal to a perturbative expansion;
rather, we will use central values and errors from the most recent direct fits
of the PMNS matrix parameters to global data.
Using the notation $[-2\sigma,{\bf{\rm best\ fit}},+2\sigma]$,
we summarize the global analysis of~\cite{Valle:2012} as
\bea{globalfitsNH}
\sin^2\theta_{13}&\subset& [\,0.019,\ \ \ \ {\bf 0.0246},\ \ \ 0.030\,]\,,\nonumber\\
\sin^2\theta_{32}&\subset& [\, 0.38,\,\,\, {\bf 0.613/0.427},\,0.66\,]\,,
\eea
if the neutrino mass-squared ordering displays a ``normal'' hierarchy (i.e., $(m_3^2-m_2^2)\gg (m_2^2-m_1^2) >0$);
and slightly different,
\bea{globalfitsIH}
\sin^2\theta_{13}&\subset& [\,0.020,{\bf 0.0250},0.030\,]\,,\nonumber\\
\sin^2\theta_{32}&\subset& [\,0.39,\ \, {\bf 0.600},\ 0.65\ ]\,,
\eea
if the hierarchy is ``inverted'' (i.e., $(m_2^2-m_3^2)\gg(m_2^2-m_1^2)>0$).
The mass ordering $m_1<m_2$ is fixed by the matter effect in the Sun needed to explain
the observed solar ratio $w_e\sim 1/3$.
This leaves the two possible hierarchical orderings identified above.
The claim is made in Ref.~\cite{Valle:2012} that $\sin^2\theta_{13}$ is $10.2\,\sigma$ away from zero.
The first and second ``best fit'' options for $\sin^2\theta_{32}$ in the normal hierarchy
reflect an octant ambiguity in the present data.
We remark that $\sin^2\theta_{32}$ is not maximal in the best fit value, nor within the $1\sigma$ error,
but may be maximal at $2\sigma$.

For either hierarchy, the remaining results of the global fit concern $\sin^2\theta_{12}$ and $\cos\delta$:
\beq{globalfit12}
\sin^2\theta_{12} \subset  [\ 0.29,{\bf 0.320},0.35\ ]\,,
\eeq
which shows that $\sin^2\theta_{12}$ is many $\sigma$ away
from either zero or the maximal value of $1/2$.
For $\cos \delta$, at even $1\sigma$ the entire range of $[ -1,+1 ]$ is allowed.
We remark that even though our propagation matrix elements $P_{\alpha\beta}$ have a classical
explanation (given in the introduction), they nevertheless depend on the
quantum mechanical ``Dirac phase'' parameter $\delta$, via the
$CP$-conserving factor $\textrm{Re} (e^{i\delta}) = \cos\delta$.
We will take three typical values $(0,+1,-1)$ for the unconstrained parameter $\cos\delta$.
The global analysis of~\cite{Fogli:2012,GonzalezGarcia:2012sz}
find numbers similar to those used here~\cite{Valle:2012}.
We note that for the choice $\cos\delta = 0$,
the deviation of $\theta_{32}$ from maximality ($\pi/4$) is the sole source of
$\numu$-$\nutau$~symmetry breaking.
(Accordingly, ${\rm Det}(P)$ will be very small for the choice $\cos\delta=0$.)

Experiments necessarily determine the above parameters in combinations.
Accordingly, the parameter errors quoted above are correlated.
We will take these errors as uncorrelated, since the alternative
requires an independent global fit for each change of any parameter~\cite{Grimus}.
Treating the errors as uncorrelated is conservative in that it allows a larger range of
parameter values for a given confidence level.
%
%

We illustrate the result of the global best fit for the normal hierarchy case, with $\theta_{32}<\pi/4$.
The three entries per matrix correspond to three assumed values for the Dirac~CP phase,
with ordering $\cos\delta=0,+1,-1$.
We find:
\begin{widetext}
\be
\label{UsqNumer}
|U_{\alpha j}|^2  =
\left(
  \begin{array}{ccc}
  0.663 & 0.312 & 0.0246 \\
  0.191 & 0.393 & 0.416 \\
  0.146 & 0.295 & 0.559
  \end{array}
\right)\,,
\left(
  \begin{array}{ccc}
  0.663 & 0.312 & 0.0246 \\
  0.263 & 0.321 & 0.416  \\
  0.0738& 0.368 & 0.559
  \end{array}
\right)\,,
\left(
  \begin{array}{ccc}
  0.663 &  0.312  &  0.0246 \\
  0.118 &  0.465  &  0.416  \\
  0.219 & 0.222   &  0.559
  \end{array}
\right)\,;
\\ \\
\eeq
and the symmetric matrices
\begin{eqnarray}
\label{PNumer}
P &=&
\left(
  \begin{array}{ccc}
  0.538   &    0.259  &   0.203 \\
   \cdots &    0.364  &   0.377 \\
   \cdots &    \cdots &   0.421
  \end{array}
\right)\,,
\left(
  \begin{array}{ccc}
  0.538   &    0.285 &   0.177 \\
   \cdots                 &    0.345  &   0.370 \\
   \cdots                 &    \cdots                  &   0.453
  \end{array}
\right)\,,
\left(
  \begin{array}{ccc}
  0.538   &  0.234  &  0.228 \\
   \cdots &  0.404  &  0.362 \\
   \cdots &  \cdots &  0.410
  \end{array}
\right)\,;
\\ \nonumber\\
\label{DetPinvNumer}
\textrm{Det}(P)\, P^{-1}
&=&
\left(
  \begin{array}{ccc}
   0.0115 &  -0.0327 &  0.0238 \\
   \cdots &   0.185  & -0.150 \\
   \cdots &  \cdots  &  0.129
  \end{array}
  \right)\,,
\left(
  \begin{array}{ccc}
   0.0195 &  -0.0633 &  0.0441 \\
   \cdots &   0.212  & -0.149  \\
   \cdots &  \cdots  &  0.105
  \end{array}
  \right)\,,
\left(
  \begin{array}{ccc}
   0.0344 &  -0.0132  & -0.00750 \\
   \cdots  &  0.168   & -0.141 \\
   \cdots  &  \cdots   &  0.163
  \end{array}
  \right)\,;
\\ \nonumber\\
\label{PinvNumer}
P^{-1} &=&
\left(
      \begin{array}{ccc}
      4.59   & -13.1   &  9.52 \\
      \cdots &  74.2   & -60.1 \\
      \cdots &  \cdots & 51.5
    \end{array}
    \right)\,,
     \left(
 \begin{array}{ccc}
      66.9     &  -217 &  151 \\
      \cdots   &  728  & -510 \\
      \cdots   & \cdots & 359
    \end{array}
    \right)\,,
\left(
      \begin{array}{ccc}
      2.51    &  -0.961  &   -0.548 \\
      \cdots  &   12.3   &  -10.3\\
      \cdots  &   \cdots &   11.9
    \end{array}
    \right)\,;
\end{eqnarray}
\end{widetext}
\beq{DetNumer}
{\rm where\ \ Det}(P) = (2.50, 0.291,13.7)\times10^{-3}\,.
\eeq
The $|U|^2$ and $P$ matrices may be compared to the corresponding TBM matrices,
given earlier in Eqs.~\rf{UsquaredTBM} and~\rf{P-TBM}.
The determinant of $P_{\rm TBM}$ vanishes, and so there is no $P_{\rm TBM}^{-1}$ to which one may compare.
On the other hand, the matrix ${\rm Det}(P)\,P^{-1}$ has the form of the analogous TBM matrix,
given in Eq.~\rf{TBMinverse}.

A visual comparison between the TBM and Nature's choices is given in Fig. 1.
The constraint $w_e+w_\mu+w_\tau=1$ reduces ${\vec w}$-space to a triangle with corners
at (1,0,0), (0,1,0), and (0,0,1).
We orient the triangle with $w_e$ at the apex.  Then, $\numu$-$\nutau$~symmetry with its $w_\mu = w_\tau$
defines a vertical line through the center of the triangle.  The horizontal distance of the point ${\vec w}$ from the
center line provides a kind of measure of $\numu$-$\nutau$~symmetry breaking.
\begin{widetext}
%
%
\begin{figure}
\captionsetup[subfigure]{labelformat=empty}
\centering
\begin{tabular}{cc}
\subfloat[(a)]
{\label{fig:triangles}
\includegraphics[width=0.24\textwidth]{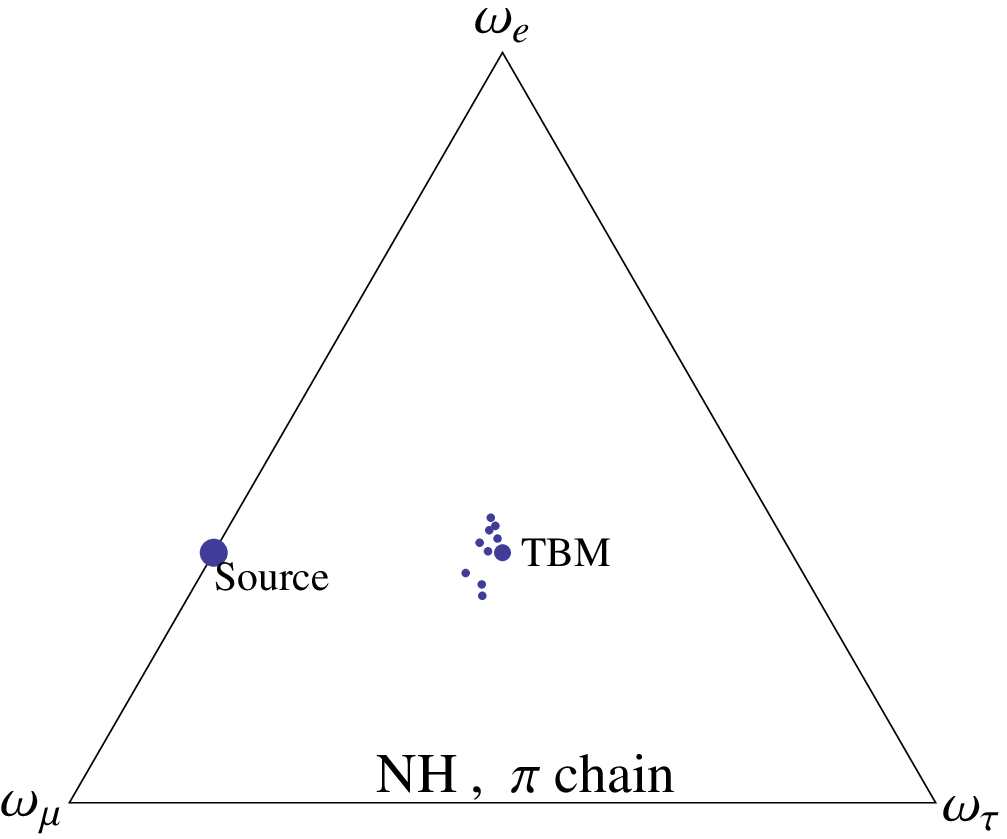}}
\subfloat[(A)]{
\includegraphics[width=0.24\textwidth]{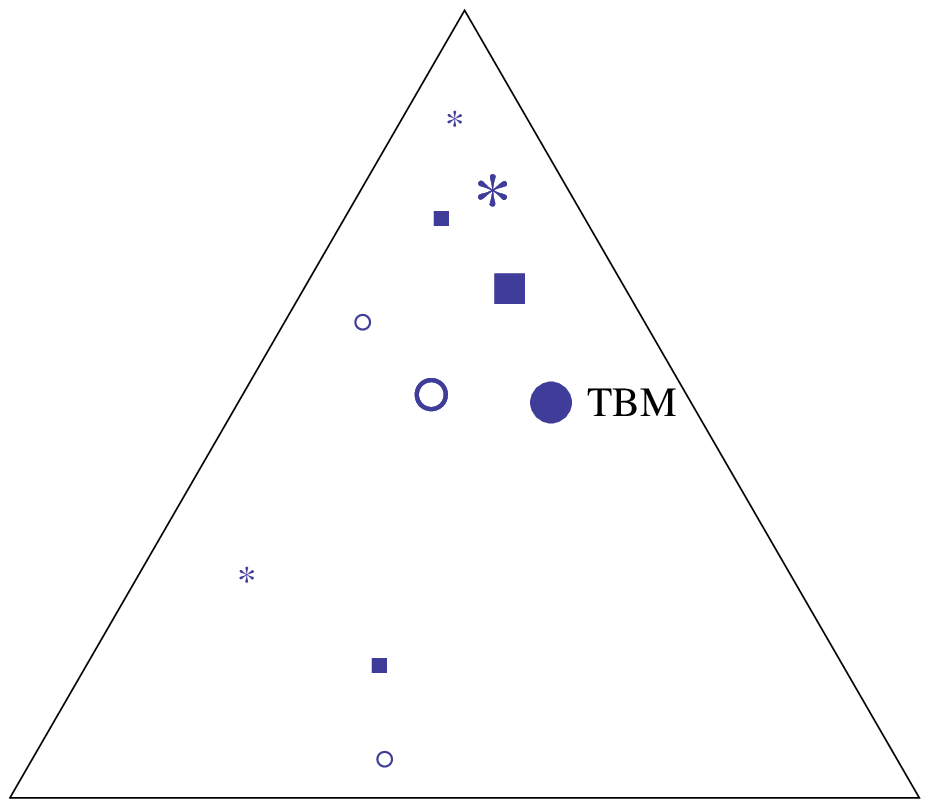}}\\
\subfloat[(b)]{
\includegraphics[width=0.24\textwidth]{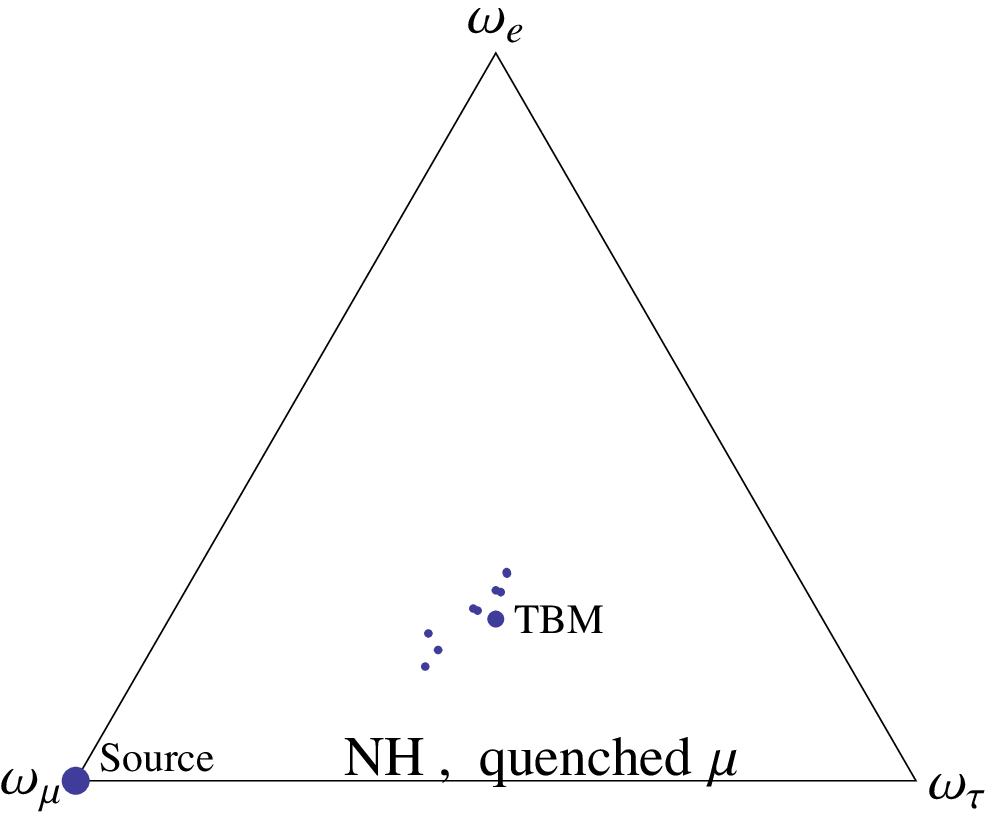}}
\subfloat[(B)]{
\includegraphics[width=0.24\textwidth]{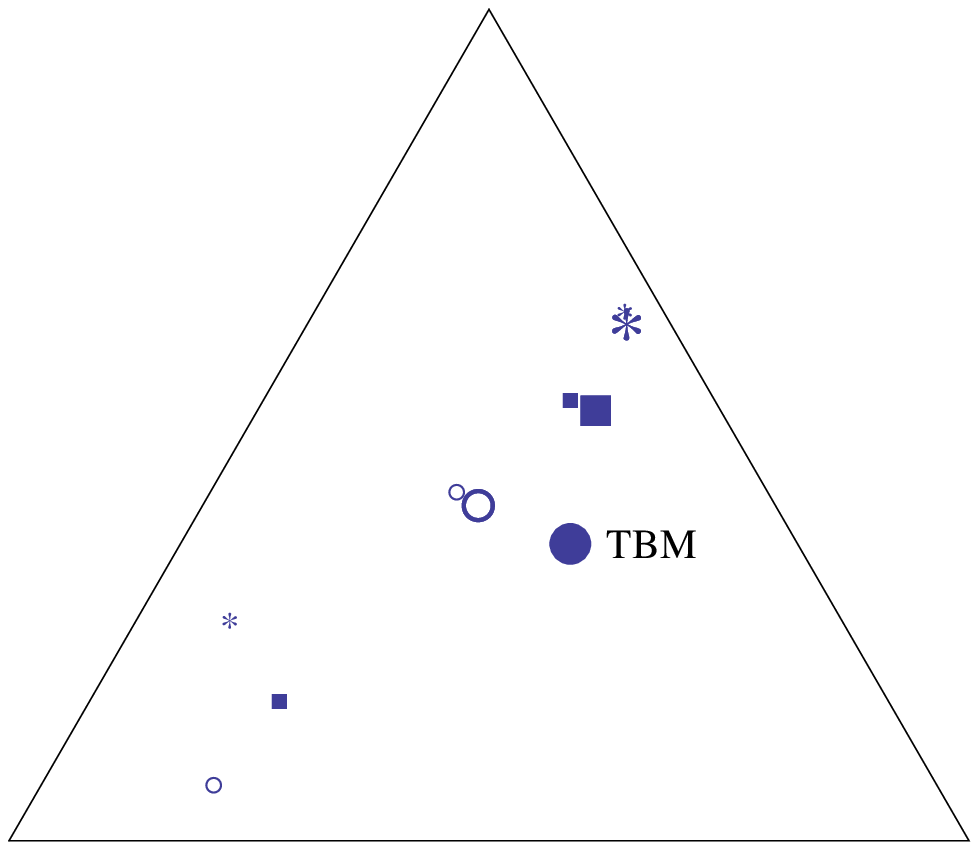}}\\
\subfloat[(c)]{
\includegraphics[width=0.24\textwidth]{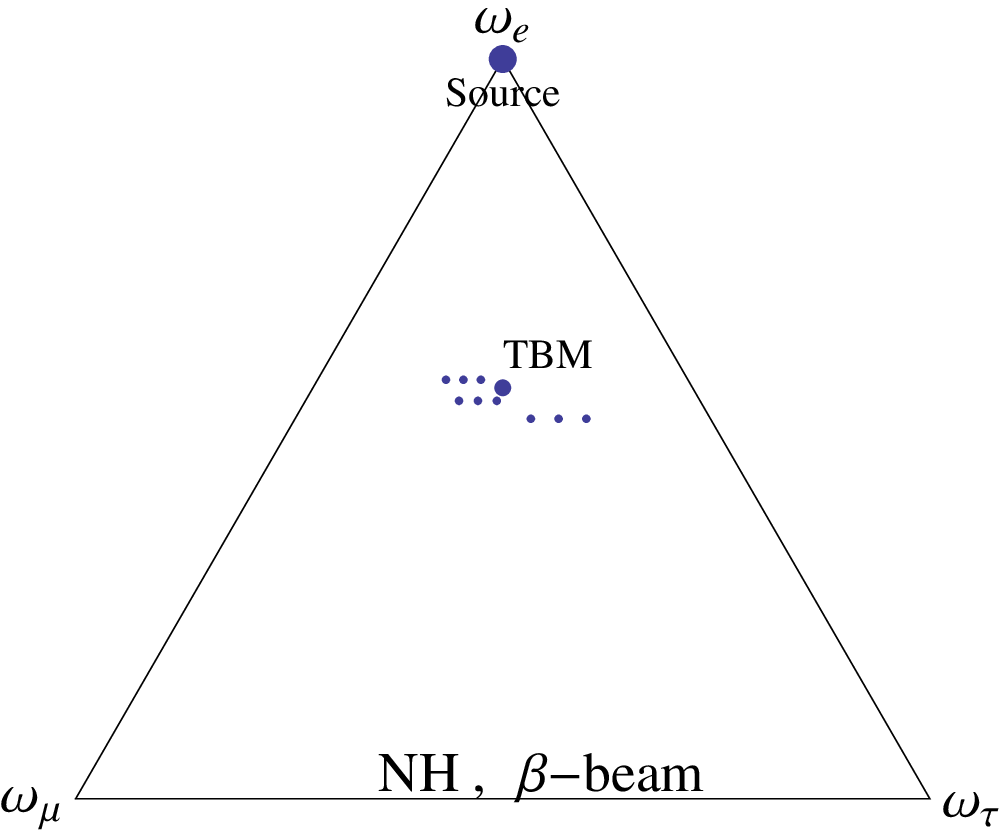}}
\subfloat[(C)]{
\includegraphics[width=0.24\textwidth]{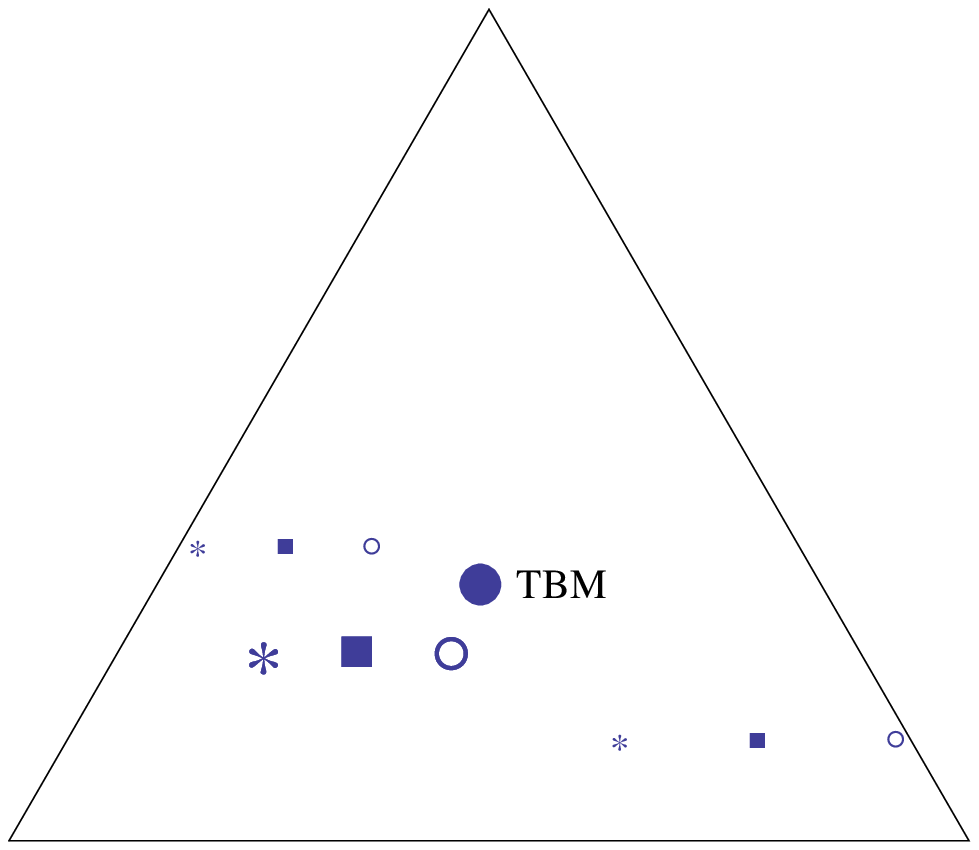}}\\
\subfloat[(d)]{
\includegraphics[width=0.24\textwidth]{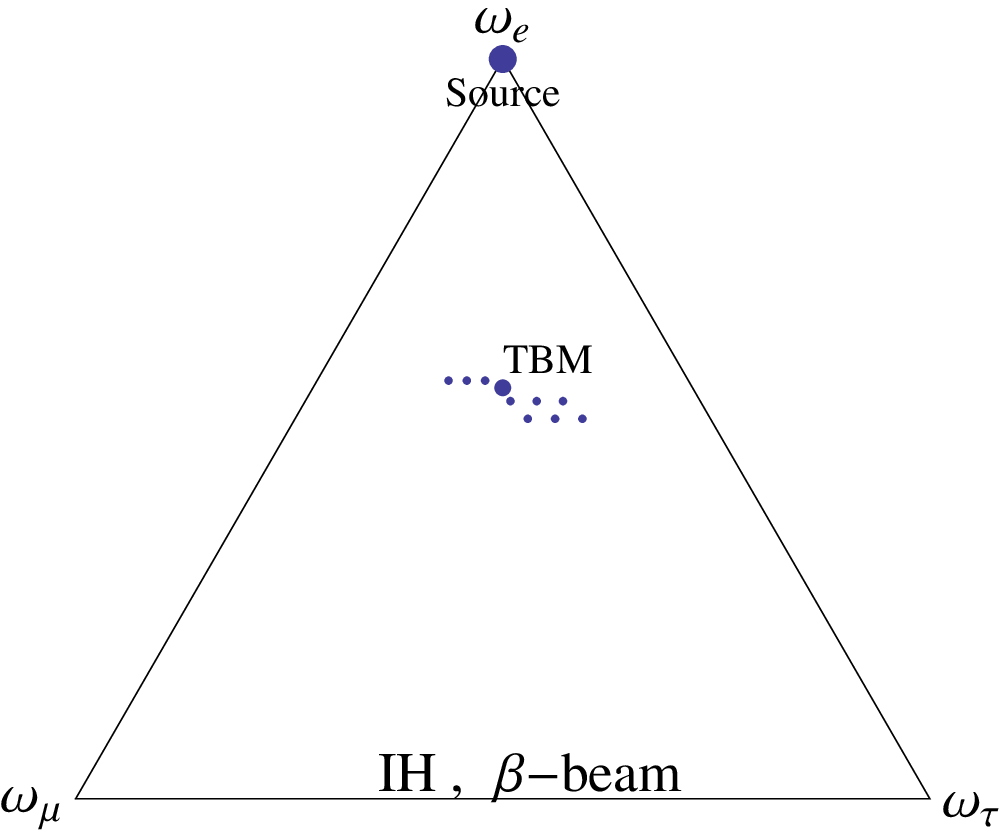}}
\subfloat[(D)]{
\includegraphics[width=0.24\textwidth]{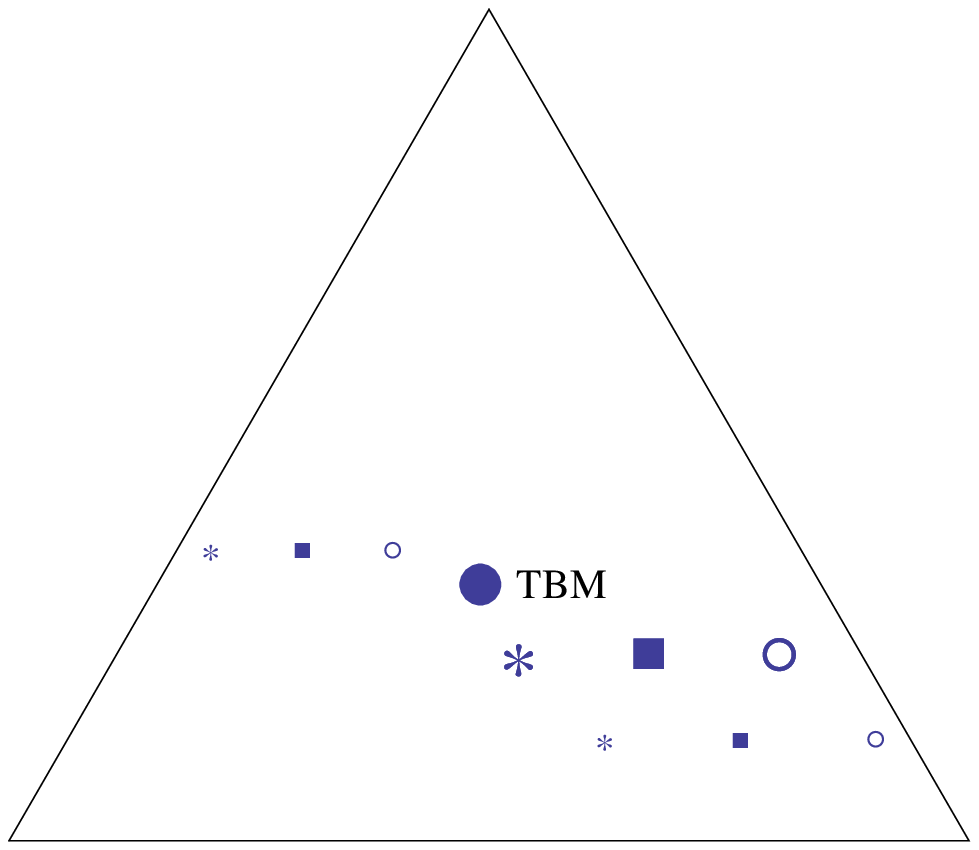}}
\end{tabular}
\caption{Triangle plots of (left) entire ${\vec w}$-parameter space,
(right) un-normalized blow-up of left panel parameter region.
See the main text for an explanation of symbols.}
\end{figure}
%
\end{widetext}

The left triangle plots (lower-case letters) show the entire ${\vec w}$-parameter space,
for three different flavor-injection models
for normal (NH) and inverted (IH) neutrino mass hierarchies.
In descending order, the plots are:
\begin{itemize}
  \item (a) NH with pion chain injection,
  \item (b) NH with quenched muon (incomplete pion chain) injection,
  \item (c) NH with $\beta$-beam injection, and
  \item (d) IH with $\beta$-beam injection.
\end{itemize}

Each plot shows the flavor values for TBM (large solid dot) and nine updated sets of fitted mixing angles:
square, star, and open circle correspond to Dirac phase $\delta=\pi/2, 0, \pi$, respectively(i.e. $\cos\delta =$~(0, 1, -1)).
Larger symbols correspond to best fit values for mixing angles and smaller symbols to $\pm2\sigma$~values.
Combinations of hierarchies and injection models not shown would appear very similar to
one of the four plots on display.
Right panels (capital letters A,B,C,D) show un-normalized blow-up of left panel parameter region
containing the nine predictions and TBM value.
$\numu$-$\nutau$ symmetry predicts a value on the vertical line through the TBM dot;
deviations from $\numu$-$\nutau$ symmetry are evident.

\subsection{Examples of Three-Flavor Phenomenology}
\label{subsec:Pheno}
Another use of the numerical results is to input the $P_{\alpha\beta}$ values from Eq.~\rf{PNumer}
into Eqs.~\rf{w-maxmin} and \rf{w-maxmin-notau}.
Here we learn, for example, that $w_e$ is bounded, in the NH model with $\theta_{32}<\pi/4$,
by a maximum of 0.538, and by a minimum of (0.177, 0.203, 0.228), for $\cos\delta=(0,+1,-1)$.
When $W_\tau$ is set to zero to conform with theoretical prejudice,
then the maximum is not affected but the minimum rises to (0.285, 0.259, 0.234), respectively.

Although the main use of the inverse propagation matrix which we have constructed is implementation of
Eq.~\rf{FlavorProp2} to infer neutrino flavor ratios at the cosmic sources,
here we present yet another use for the matrix $P^{-1}$.
Compelling theoretical arguments from particle physics tell us that $\nutau$ production at the source
is very suppressed, due to the heavy $\tau$~mass.  Thus we expect $W_\tau\sim 0$.
This expectation can be easily checked.
From Eq.~\rf{FlavorProp2} we have
\beq{Wtau1}
0= W_\tau = P_{e\tau}^{-1}w_e + P_{\mu\tau}^{-1}w_\mu +P_{\tau\tau}^{-1}w_\tau\,,
\eeq
with the elements of $P^{-1}$ given analytically in Eq.~\rf{Pinverse1}.
Multiplied by ${\rm Det}(P)$, this result is
\bea{Wtau2}
0 &=& (P_{e\mu}P_{\mu\tau} - P_{e\tau}P_{\mu\mu})\,w_e \\
   &+& (P_{e\tau}P_{e\mu}-P_{ee}P_{\mu\tau})\,w_\mu + (P_{ee}P_{\mu\mu}-P_{e\mu}^2)\,w_\tau\,,\nonumber
\eea
subjecting to the normalization $w_e+w_\mu+w_\tau=1$.
Numerical values for the parenthetical expressions for the best fit of the normal hierarchy
with $\theta_{32}$ in the first octant are given in the final columns of Eq.~\rf{DetPinvNumer}.
Any observed violation of this result would implicate $\nutau$ production at the sources.

As a final demonstration of the utility of the new, full flavor evolution matrix,
we consider the dependence of the flavor ratio $W_\mu/W_e$ at the sources on the flavor ratios
observed at Earth.
Here, we embrace theoretical prejudice and assume that $\nutau$'s are not produced at the sources, i.e. $W_\tau=0$.
We use Eq.~\rf{FlavorProp1} to derive the three $w_\alpha (W_e,W_\mu)$, then sum
$w_e$ and $w_\tau$ to get $w_{sh}$, from the Earthly ratio
$w_\mu(W_e,W_\mu)/w_{sh}(W_e,W_\mu)$, and invert to get
\beq{ratioN}
\frac{W_\mu}{W_e} = \frac{P_{e\mu}-(P_{ee}+P_{e\tau})\left(\frac{w_\mu}{w_{sh}}\right)}
                         {(P_{e\mu}+P_{\mu\tau})\left(\frac{w_\mu}{w_{sh}}\right) - P_{\mu\mu}}\,.
\eeq
This equation generalizes Eq.~\rf{R(r)2} to the condition of broken $\numu$-$\nutau$ symmetry.
It is independent of any injection model.
Therefore, it allows us to infer the flavor ratio of cosmically distant sources
from the track-to-shower ratio observed at Earth, and thereby discriminate among injection models.
Values for the injection ratio expected from the most popular source models are
$\sim (2,\infty,0)$ for the pion decay-chain, quenched muon, and $\beta$-beam models, respectively.
These values for the injection ratio are quite different,
and so discrimination among popular models via Eq.~\rf{ratioN} should be straightforward.

\section{Discussions and Conclusions}
\label{sec:Conclusions}
Experimental inference of flavor ratios involves some uncertainty.
We have neglected these uncertainties in this more theoretical paper.
For example, neutrino neutral-current (NC) interactions, the same for all three flavors,
contribute to shower events.
The ratio of the neutral to charged current cross-sections is known, so the NC
contribution can be accounted for in a data sample.
As another example, the experimental efficiencies for measuring shower events and muon-track events are different.
Again, these can be accounted for in a data sample.

What we have shown is that neutrino flavor physics is rich in its information content,
and therefore worth pursuing.
Neutrino flavor physics offers another window into the dynamics of the most distant, most energetic
objects in the Universe.

We end with a summary of the main results discussed in this paper.  We have:
\begin{itemize}
  \item shown the movement of Earthly measured flavor ratios away from the $\numu$-$\nutau$ symmetric value
of the previously viable TriBiMaximal model, for the three most popular cosmic-source flavor models;

  \item derived the inverse flavor propagation matrix which allows one to infer flavor ratios injected at
cosmically-distant sources from the ratios observed here on Earth;

  \item presented a relation among flavor ratios at Earth that determines whether $\nutau$'s are injected at the source;

  \item derived a general formula for the $\numu/\nue$ injection flavor ratio at the source in terms of the
observable ratio of track-to-shower events at Earth.
\end{itemize}

\vspace{1cm}

Acknowledgements:
This research was supported in part by Department of Energy grant DE-FG05-85ER40226.

\end{document}